\documentclass[12pt]{article}
\usepackage{amsmath}
\usepackage{bm}
\usepackage{color}

\begin{document}
\begin{center}
\begin{large}
{\bf Minimal length estimation on the basis of studies of the Sun-Earth-Moon system in deformed space}
\end{large}
\end{center}

\centerline {Kh. P. Gnatenko \footnote{E-Mail address: khrystyna.gnatenko@gmail.com},  V. M. Tkachuk \footnote{E-Mail address: voltkachuk@gmail.com}}
\medskip
\centerline {\small \it Ivan Franko National University of Lviv, Department for Theoretical Physics,}
\centerline {\small \it 12 Drahomanov St., Lviv, 79005, Ukraine}

\begin{abstract}
A space with deformed Poisson brackets for coordinates and momenta leading to the minimal length is considered. Features of description of motion of a body in the space are examined. We propose conditions on the parameters of deformation on which Poisson brackets for coordinates and momenta of the center-of-mass reproduce relations of deformed algebra, kinetic energy of a body is independent of its composition, and the weak equivalence principle is preserved in the deformed space. Influence of minimal length on the motion of the Sun-Earth-Moon system is studied. We find that deformation of the Poisson brackets leads to corrections to the accelerations of the Earth and the Moon toward the Sun, as a result the Eotvos-parameter does not vanish even if we consider equality of gravitational and inertial masses. The upper bound for the minimal length is estimated using results of the Lunar laser ranging experiment.
\end{abstract}


\section{Introduction}

Studies of a space with deformed commutation relations for coordinates and momenta leading to the minimal length have attracted much attention owing to development of string theory and quantum gravity (see, for example, \cite{Witten,Witten1,Maggiore,Doplicher}).

Deformation of commutation relations for coordinates and momenta in one dimensional case
\begin{eqnarray}
[X,P]=i\hbar(1+\beta P^2),{}\label{d}
\end{eqnarray}
with $\beta$ being the parameter of deformation leads to the minimal length  $\hbar\sqrt{\beta}$ \cite{Kempf95,Kempf96}. In the cases of higher dimensions deformed algebra (\ref{d}) can be generalized as
 \begin{eqnarray}
[X_{i},X_{j}]=i\hbar\frac{(2\beta-\beta^{\prime})+(2\beta+\beta^{\prime})\beta P^2}{1+\beta P^2}(P_iX_j-P_jX_i),{}\label{0al}\\{}
[X_i,P_j]=i\hbar(\delta_{ij}(1+\beta P^2)+\beta^{\prime}P_iP_j),{}\label{0al1}\\{}
[P_{i},P_{j}]=0,\label{0al2}
 \end{eqnarray}
here $\beta\geq0$, $\beta^{\prime}\geq0$ are parameters of deformation  \cite{Kempf97,Benczik02,Benczik,Tkachuk07}.  In the space (\ref{0al})-(\ref{0al2})  the minimal length is determined as $\hbar\sqrt{\beta+\beta^{\prime}}$.
In particular case $\beta=0$ from  (\ref{0al})-(\ref{0al2}) one obtains relations corresponding to the nonrelativistic Snyder model (see, for instance, \cite{Mignemi11,Mignemi12,Mignemi15})
 \begin{eqnarray}
[X_{i},X_{j}]=i\hbar\beta^{\prime}(P_jX_i-P_iX_j),{}\label{Sal}\\{}
[X_i,P_j]=i\hbar(\delta_{ij}+\beta^{\prime}P_iP_j),{}\label{Sal1}\\{}
[P_{i},P_{j}]=0.\label{Sal2}
 \end{eqnarray}
If $\beta^{\prime}=2\beta$ from (\ref{0al})-(\ref{0al2})  in the first order in parameters of deformation one obtains deformed algebra with commuting coordinates  and commuting momenta \cite{Brau,Tk2}
 \begin{eqnarray}
[X_{i},X_{j}]=[P_{i},P_{j}]=0,\label{a1}{}\\{}
[X_i,P_j]=i\hbar(\delta_{ij}(1+\beta P^2)+2\beta P_iP_j).{}\label{a2}
 \end{eqnarray}

In the present paper we consider features of description of motion of a body in a space with minimal length (\ref{0al})-(\ref{0al2}). The studies are important for examining the effect of minimal length on a vide class of physical systems. In \cite{Tk2} the authors studied the motion of a composite system (a body) in the frame of deformed algebra (\ref{0al})-(\ref{0al2}) in the spacial case of $\beta^{\prime}=2\beta$ which leads to relations (\ref{a1})-(\ref{a2}). In \cite{Gnatenkoarxiv19} a macroscopic body was examined in the Snyder space ($\beta=0$,  (\ref{Sal})-(\ref{Sal2})). In the present paper we study composite system (macroscopic body) in the frame of algebra  (\ref{0al})-(\ref{0al2}) in a general case (the  parameters of deformation $\beta$, $\beta^{\prime}$ are positive constants considered to be different for different particles). We examine  relations of deformed algebra for coordinates and momenta of the center-of-mass of a body.  The properties of the kinetic energy of a body in the deformed space are studied. We conclude that the idea to relate parameters of deformation corresponding to a particle with its mass proposed in \cite{Tk2,Gnatenkoarxiv19} is important also in the case of deformed algebra (\ref{0al})-(\ref{0al2}). This idea gives a possibility to recover independence of the kinetic energy of composition in the space (\ref{0al})-(\ref{0al2}), to consider deformed algebra (\ref{0al})-(\ref{0al2}) with effective parameters of deformation for description of the center-of-mass motion.

Deformation of commutation relations  for coordinates and momenta opens a possibility to describe
quantum nature of space on a phenomenological level. At the same time the deformation leads to the fundamental problems, among them the problem of violation of the weak equivalence principle.
This problem has been studied in the frame of one dimensional deformed algebra \cite{Tk1}, Snyder algebra  \cite{Mignemi14,Gnatenkoarxiv19}, noncommutative algebra of canonical type \cite{Bastos1,GnatenkoPLA13,Saha,GnatenkoPLA17,Bertolami2,Saha1,GnatenkoEPL}, noncommutative algebra of Lie-type \cite{GnatenkoPRD}. It is important to note  that as was concluded in  \cite{Tk1,GnatenkoPLA13,GnatenkoPLA17,GnatenkoEPL,GnatenkoPRD,Gnatenkoarxiv19} idea to relate parameters of deformed algebras with mass opens a possibility to recover the weak equivalence principle in quantum space.

In the present paper we consider the weak equivalence principle in the frame of deformed algebra (\ref{0al})-(\ref{0al2}).  We study effect of minimal length on the motion of the Sun-Earth-Moon system and find corrections to the Eotvos-parameter for the Earth and the Moon in the space (\ref{0al})-(\ref{0al2}). On the basis of the studies the minimal length is estimated.

The paper is organized as follows. In Section 2 we consider the problem of description of motion of a body in a space with minimal length (\ref{0al})-(\ref{0al2}).  Section 3 is devoted to examining of the weak equivalence principle in the deformed space with minimal length. The Sun-Earth-Moon system is studied in the frame of algebra (\ref{0al})-(\ref{0al2}) and the upper bound for the minimal length is estimated.
 Conclusions are presented in Section 4.

\section{Motion of a body in deformed space with minimal length and parameters of deformation}

Let us consider features of description of motion of a body in deformed space with minimal length (\ref{0al})-(\ref{0al2}). We start with the following Hamiltonian
\begin{eqnarray}
H=\frac{P^2}{2M},\label{h}
\end{eqnarray}
here $M$ is the mass of the body, $P^2=\sum_iP_i^2$, momenta $P_i$ satisfy relations of deformed algebra (\ref{0al})-(\ref{0al2}) which in the classical limit read
\begin{eqnarray}
\{X_{i},X_{j}\}=\frac{(2\beta-\beta^{\prime})+(2\beta+\beta^{\prime})\beta P^2}{1+\beta P^2}(P_iX_j-P_jX_i),{}\label{pois1}\\{}
\{X_i,P_j\}=\delta_{ij}(1+\beta P^2)+\beta^{\prime}P_iP_j,{}\label{pois2}\\{}
\{P_{i},P_{j}\}=0.\label{pois3}
 \end{eqnarray}
Taking into account  Poisson brackets (\ref{pois1})-(\ref{pois3}) one obtains the following equations of motion
\begin{eqnarray}
\dot{X}_i=\frac{P_i}{M}(1+(\beta+\beta^{\prime})P^2),\label{eq1}\\
\dot{P}_i=0.
\end{eqnarray}
From (\ref{eq1}) one has that the relation between momenta and velocities is deformed.  Up to the first order in $\beta$ and $\beta^{\prime}$ one can write
\begin{eqnarray}
P_i=\frac{M\dot{X}_i}{1+(\beta+\beta^{\prime})M^2\dot{X}^2},\label{rel}
\end{eqnarray}
here $\dot X^2=\sum_i\dot X_i^2$. Using (\ref{rel}), one has
\begin{eqnarray}
H=\frac{M\dot{X}^2}{2}(1-2(\beta+\beta^{\prime})M^2\dot{X}^2).\label{hh}
\end{eqnarray}
On the other hand one can consider macroscopic body as a composite system made of $N$ particles. Let us divide the body into $N$ parts of masses $m_a$ which move with the same velocities as the whole body and can be considered as particles. So, let us study a composite system made of $N$ particles in deformed space with Hamiltonian
\begin{eqnarray}
H=\sum_a\frac{(P^{(a)})^2}{2m_a}.\label{hadd}
\end{eqnarray}
Index $a$ is used to label the particles, $a=(1..N)$.

 We assume that Poisson brackets for coordinates and momenta of different particles vanish and consider a general case when coordinates and momenta of different particles satisfy the deformed algebra with different parameters
\begin{eqnarray}
\{X^{(a)}_{i},X^{(b)}_{j}\}=\delta_{ab}\frac{(2\beta_a-\beta_a^{\prime})+(2\beta_a+\beta_a^{\prime})\beta_a (P^{(a)})^2}{1+\beta_a (P^{(a)})^2}(P^{(a)}_iX^{(a)}_j-P^{(a)}_jX^{(a)}_i),{}\label{ppoisl}\\{}
\{X^{(a)}_i,P^{(b)}_j\}=\delta_{ab}\delta_{ij}(1+\beta_a (P^{(a)})^2)+\delta_{ab}\beta_a^{\prime}P^{(a)}_iP^{(a)}_j,{}\label{ppois2}\\{}
\{P^{(a)}_{i},P^{(a)}_{j}\}=0,\label{ppois3}
 \end{eqnarray}
 here indexes $a$ and $b$ label the particles.
 So, taking into account (\ref{ppoisl})-(\ref{ppois3}), one obtains the following equations of motion
 \begin{eqnarray}
\dot{X}^{(a)}_i=\frac{P^{(a)}_i}{m_a}(1+(\beta_a+\beta_a^{\prime})(P^{(a)})^2),\label{eqp1}\\
\dot{P}^{(a)}_i=0.
\end{eqnarray}
Using (\ref{eqp1}), one can rewrite Hamiltonian (\ref{hadd}) up to the first order in the deformation parameters as follows
\begin{eqnarray}
H=\frac{M\dot{X}^2}{2}(1-2M^2(\tilde\beta+\tilde\beta^{\prime})\dot{X}^2),\label{hadd1}
\end{eqnarray}
where $M=\sum_am_a$. We use notations
\begin{eqnarray}
\tilde\beta=\sum_a\beta_a\mu^3_a, \label{beff}\\
\tilde\beta^{\prime}=\sum_a\beta_a^{\prime}\mu^3_a, \label{beff1}
\end{eqnarray}
and take into account that particles forming the body move with velocities which are equal to the velocity of the body
\begin{eqnarray}
\dot{X}^{(a)}_i=\dot{X}_i.
\end{eqnarray}
Note that the expressions (\ref{hh}), (\ref{hadd1}) coincide if parameters of deformation  which describe the motion of the body are defined as (\ref{beff}), (\ref{beff1}), else the property of additivity of the kinetic energy is not preserved in deformed space.

From (\ref{hadd1}) one has that the kinetic energy of the body depends on parameters $\tilde\beta$,  $\tilde\beta^{\prime}$ given by (\ref{beff}), (\ref{beff1}) therefore it depends on the  composition of the body.
 The property of independence of the kinetic energy of composition is preserved if  the following relations hold
 \begin{eqnarray}
\sqrt{\beta_a} m_a=\gamma=const,\label{c1}\\
\sqrt{\beta^{\prime}_a} m_a=\gamma^{\prime}=const,\label{c2}
\end{eqnarray}
where parameters $\beta_a$, $\beta_a^{\prime}$ correspond to a particle with mass $m_a$ and $\gamma$, $\gamma^{\prime}$ are constants which are the same for different particles. Taking into account (\ref{c1}), (\ref{c2}) from (\ref{beff}), (\ref{beff1}) one obtains
 \begin{eqnarray}
\tilde\beta=\frac{\gamma^2}{M^2},\label{efc1}\\
\tilde\beta^{\prime}=\frac{(\gamma^{\prime})^2}{M^2}.\label{efc2}
\end{eqnarray}
Parameters $\tilde\beta$, $\tilde\beta^{\prime}$ are determined by the total mass $M$ and do not depend on composition of the body. So, if conditions (\ref{c1}) and (\ref{c2}) are satisfied the property of independence of the kinetic energy on the composition is recovered in the deformed space (\ref{pois1})-(\ref{pois3}). It is worth noting that taking into account (\ref{efc1}), (\ref{efc2}), conditions (\ref{c1}), (\ref{c2}) can be generalized including effective parameters
 \begin{eqnarray}
\sqrt{\beta_a} m_a=\sqrt{\tilde{\beta}}M=\gamma=const,\label{cc1}\\
\sqrt{\beta^{\prime}_a} m_a=\sqrt{\tilde{\beta}^{\prime}} M=\gamma^{\prime}=const.\label{cc2}
\end{eqnarray}

We would like to mention that relations (\ref{c1}), (\ref{c2})  are in the agreement to that proposed to solve the list of problems in deformed space with minimal length  (\ref{a1})-(\ref{a2}) \cite{Tk1,Tk3} and in the Snyder space (\ref{Sal})-(\ref{Sal2})) \cite{Gnatenkoarxiv19}.

Another important result which can be obtained due to assumptions (\ref{c1}), (\ref{c2}) is recovering of relations of deformed algebra (\ref{pois1})-(\ref{pois3}) for coordinates and momenta of the center-of-mass.
To show that let us first analyze expression (\ref{eqp1}) in the case when conditions (\ref{c1}), (\ref{c2}) hold. From (\ref{eqp1}), (\ref{c1}), (\ref{c2}) we have that for particles which move with the same velocities $\dot{X}^{(a)}_i=\dot{X}_i$ we can write
\begin{eqnarray}
P^{(a)\prime}_i\left(1+(\gamma^2+(\gamma^{\prime})^2)({P^{(a)\prime}_i})^2\right)=\dot{X}_i,\label{eqp11}
\end{eqnarray}
where
\begin{eqnarray}
P^{(a)\prime}_i=\frac{P^{(a)}_i}{m_a}.
\end{eqnarray}
So, momenta $P^{(a)\prime}_i$ depend on the velocities $\dot{X}_i$ and  parameters $\gamma$, $\gamma^{\prime}$ which are the same for particles forming the body. Therefore, values $P^{(a)\prime}_i$ are the same for particles forming the body as it is in the ordinary space ($\beta_a=\beta_a^{\prime}=0$). On the basis of this conclusion if relations (\ref{c1}), (\ref{c2}) hold, for coordinates and momenta of the center-of-mass of a body $\tilde{\bf X}=\sum_a\mu_a{\bf X}^{(a)}$, $\tilde{\bf P}=\sum_a{\bf P}^{(a)}$ (here $X^{(a)}_i$, $P^{(a)}_i$ satisfy (\ref{ppoisl})-(\ref{ppois3})) one obtains the following Poisson brackets
\begin{eqnarray}
\{\tilde{X}_i,\tilde{X}_j\}=\nonumber\\=\sum_a\mu^2_a\frac{(2\beta_a-\beta_a^{\prime})+(2\beta_a+\beta_a^{\prime})\beta_a (P^{(a)})^2}{1+\beta_a (P^{(a)})^2}(P^{(a)}_iX^{(a)}_j-P^{(a)}_jX^{(a)}_i)=\nonumber\\=\frac{(2\tilde{\beta}-\tilde{\beta}^{\prime})+(2\tilde{\beta}+\tilde{\beta}^{\prime})\tilde{\beta} \tilde{P}^2}{1+\tilde{\beta} \tilde{P}^2}(\tilde{P}_i\tilde{X}_j-\tilde{P}_j\tilde{X}_i),\label{cmcc}\\
\{\tilde{X}_i,\tilde{P}_j\}=\sum_a\mu_a\delta_{ij}(1+\beta_a (P^{(a)})^2)+\sum_a\mu_a\beta_a^{\prime}P^{(a)}_iP^{(a)}_j=\nonumber\\=\delta_{ij}(1+\tilde{\beta} \tilde{P}^2)+\tilde{\beta}^{\prime}\tilde{P}_i\tilde{P}_j,
\label{cmcc1}\\
\{\tilde{P}_i,\tilde{P}_j\}=0,\label{cmcc2}
\end{eqnarray}
where parameters $\tilde{\beta}$ and $\tilde{\beta}^{\prime}$ are given by (\ref{efc1}), (\ref{efc2}).
To calculate  (\ref{cmcc}), (\ref{cmcc1}) we  use relations (\ref{cc1}), (\ref{cc2}) and take into account that if $P^{(a)\prime}_i$ are the same for different particles the following relation is satisfied $P^{(a)}_i=m_a\tilde P_i/M$, here $M$ is the total mass.

So, we can conclude that if conditions (\ref{c1}), (\ref{c2}) are satisfied the Poisson brackets for coordinates and momenta of the center-of-mass of a body (\ref{cmcc})-(\ref{cmcc2}) reproduce  relations of the deformed algebra (\ref{ppoisl})-(\ref{ppois3}) with effective parameters of deformation (\ref{efc1}), (\ref{efc2}).

\section{Sun-Earth-Moon system in deformed space and the weak equivalence principle}

Deformation of Poisson brackets for coordinates and momenta (\ref{pois1})-(\ref{pois3}) leads to violation of the weak equivalence principle. For a particle of mass $m$ in gravitational field $V(X_1,X_2,X_3)$ described by Hamiltonian
\begin{eqnarray}
H=\frac{{P}^2}{2m}+mV(X_1,X_2,X_3),\label{hh12}
\end{eqnarray}
taking into account relations (\ref{pois1})-(\ref{pois3}), one can find the following equations of motion
\begin{eqnarray}
\dot X_i=\frac{P_i}{m}\left(1+(\beta+\beta^{\prime})P^2\right)+\nonumber\\+m\frac{(2\beta-\beta^{\prime})+(2\beta+\beta^{\prime})\beta P^2}{1+\beta P^2}(P_iX_j-P_jX_i)\frac{\partial V}{\partial X_j},\label{v1}\\
\dot P_i=-m\left(1+\beta P^2\right)\frac{\partial V}{\partial X_i}-m\beta^{\prime}P_i P_j\frac{\partial V}{\partial X_j}.\label{v2}
\end{eqnarray}
 Note that because of deformation the motion of a particle in gravitational field depends on its mass therefore the weak equivalence principle is violated, even if we consider the inertial mass  (see the first term in (\ref{hh12})) to be equal to the gravitational mass (see the second term in (\ref{hh12})). We would like to stress that if  conditions on the parameters of deformation  (\ref{c1}), (\ref{c2}) are satisfied we can rewrite equations (\ref{v1}), (\ref{v2}) as
\begin{eqnarray}
\dot X_i={P^{\prime}_i}\left(1+(\gamma+\gamma^{\prime})(P^{\prime})^2\right)+\nonumber\\+\frac{(2\gamma-\gamma^{\prime})+(2\gamma+\gamma^{\prime})\gamma (P^{\prime})^2}{1+\gamma (P^{\prime})^2}(P^{\prime}_iX_j-P^{\prime}_jX_i)\frac{\partial V}{\partial X_j},\label{vv1}\\
\dot P^{\prime}_i=-\left(1+\gamma (P^{\prime})^2\right)\frac{\partial V}{\partial X_i}-\gamma^{\prime}P^{\prime}_i P^{\prime}_j\frac{\partial V}{\partial X_j},\label{vv2}
\end{eqnarray}
here $P^{\prime}_i=P_i/m$. Equations (\ref{vv1}), (\ref{vv2}) do not contain the mass $m$, therefore  the solutions $X_i(t)$, $P^{\prime}_i(t)$ also do not depend on the mass. As a result particles with different masses move in gravitational field on the same trajectories. So, the weak equivalence principle which states that the trajectory of  a particle in gravitational field is independent of its mass and composition is preserved due to the relations (\ref{c1}), (\ref{c2}).

This conclusion can be generalized for the case of motion of a body of mass $M$ in the gravitational field. If relations (\ref{c1}), (\ref{c2}) hold, considering the Hamiltonian $H=\tilde{P}^2/2M+MV(\tilde{X}_1,\tilde{X}_2,\tilde{X}_3)$ with coordinates and momenta of the center-of-mass satisfying Poisson brackets (\ref{cmcc})-(\ref{cmcc2}) we obtain the following equations of motion
\begin{eqnarray}
\dot {\tilde{X}}_i={\tilde{P}^{\prime}_i}\left(1+(\gamma+\gamma^{\prime})(\tilde{P}^{\prime})^2\right)+\nonumber\\+\frac{(2\gamma-\gamma^{\prime})+(2\gamma+\gamma^{\prime})\gamma (\tilde{P}^{\prime})^2}{1+\gamma (\tilde{P}^{\prime})^2}(\tilde{P}^{\prime}_i\tilde{X}_j-\tilde{P}^{\prime}_j\tilde{X}_i)\frac{\partial V}{\partial \tilde{X}_j},\label{vvv1}\\
\dot{\tilde{P}}^{\prime}_i=-\left(1+\gamma (\tilde{P}^{\prime})^2\right)\frac{\partial V}{\partial \tilde{X}_i}-\gamma^{\prime}\tilde{P}^{\prime}_i \tilde{P}^{\prime}_j\frac{\partial V}{\partial \tilde{X}_j},\label{vvv2}
\end{eqnarray}
 where $\tilde{P}^{\prime}_i=\tilde{P}_i/M$. Solutions of equations (\ref{vvv1})-(\ref{vvv2}) do not depend on the mass of the body and its composition. So, the weak equivalence principle is recovered due to conditions (\ref{c1}), (\ref{c2}).

Let us consider the motion of the Moon and the Earth in the gravitational field of the Sun in the frame of deformed algebra (\ref{pois1})-(\ref{pois3}).  For this purpose we examine the following Hamiltonian
\begin{eqnarray}
 H=\frac{({\bf P}^E)^{2}}{2m_E}+\frac{({\bf P}^M)^{2}}{2m_M}-G\frac{m_E m_S}{R_{ES}}-G\frac{m_M m_S}{R_{MS}}-G\frac{m_M m_E}{R_{EM}},
 \label{form12.5}
 \end{eqnarray}
 with $G$ being the gravitational constant, $m_S$ is the mass of the Sun, $m_E$, $m_M$  are the masses of the Earth and the  Moon, $R_{ES}$, $R_{EM}$, $R_{MS}$ are the distances between the Earth and the Sun, the Earth and the Moon, the  Moon and the Sun, respectively. We consider the Sun to be at the origin of the coordinate system, therefore we can write
$R_{ES}=\sqrt{\sum_i(X^E_i)^2}$, $R_{MS}=\sqrt{\sum_i(X^M_i)^2}$,
$R_{EM}=\sqrt{\sum_i(X^E_i-X^M_i)^2}$,  where $X^E_i$ and  $X^M_i$ are coordinates of the center-of-mass of the Earth and the Moon.

 Taking into account conclusions presented in the previous section, for coordinates and momenta of the center-of-mass of the Earth and the Moon one can write relations of deformed algebra (\ref{ppoisl})-(\ref{ppois3})
\begin{eqnarray}
\{X^{E}_{i},X^{E}_{j}\}=\frac{(2\beta_E-\beta_E^{\prime})+(2\beta_E+\beta_E^{\prime})\beta_E (P^{E})^2}{1+\beta_E (P^{E})^2}(P^{E}_iX^{E}_j-P^{E}_jX^{E}_i),{}\label{pE1}\\{}
\{X^{E}_i,P^{E}_j\}=\delta_{ij}(1+\beta_E (P^{E})^2)+\beta_E^{\prime}P^{E}_iP^{E}_j,{}\label{puE2}\\{}
\{X^{M}_{i},X^{M}_{j}\}=\frac{(2\beta_M-\beta_M^{\prime})+(2\beta_M+\beta_M^{\prime})\beta_M (P^{M})^2}{1+\beta_M (P^{M})^2}(P^{M}_iX^{M}_j-P^{M}_jX^{M}_i),{}\label{puE1}\\{}
\{X^{M}_i,P^{M}_j\}=\delta_{ij}(1+\beta_M (P^{M})^2)+\beta_M^{\prime}P^{M}_iP^{M}_j,{}\label{ppE2}\\{}
\{P^{E}_{i},P^{E}_{j}\}=\{P^{M}_{i},P^{M}_{j}\}=\{X^{E}_{i},P^{M}_{j}\}=\nonumber\\=\{X^{M}_{i},P^{E}_{j}\}=\{X^{M}_{i},X^{E}_{j}\}=\{P^{M}_{i},P^{E}_{j}\}=0,\label{pE3}
 \end{eqnarray}
where $\beta_E$,  $\beta_E^{\prime}$, $\beta_M$,  $\beta_M^{\prime}$  are parameters of deformation corresponding to motion of the Moon and the Earth.
Considering (\ref{form12.5}) with relations (\ref{pE1})-(\ref{pE3}) one finds the following equations of motion for the Earth
\begin{eqnarray}
\dot{{\bf X}}^E=\frac{{\bf P}^E}{m_E}\left(1+(\beta_E+\beta_E^{\prime})(P^{E})^2\right)-\frac{(2\beta_E-\beta_E^{\prime})+(2\beta_E+\beta_E^{\prime})\beta_E (P^{E})^2}{1+\beta_E (P^{E})^2}\times\nonumber\\\times\left(\frac{Gm_Em_S [{\bf X}^E\times{\bf J}^E]}{R_{ES}^{3}}+\frac{Gm_Em_M[({\bf X}^E-{\bf X}^M)\times{\bf J}^E]}{R_{EM}^{3}}\right),\label{E020}\\
\dot{{\bf P}}^E=-\frac{Gm_Em_S}{R_{ES}^{3}}\left((1+\beta_E (P^{E})^2){\bf X}^E+\beta_E^{\prime}({\bf X}^E\cdot{\bf P}^E){\bf P}^E\right)-\nonumber\\-\frac{Gm_Em_M}{R_{EM}^{3}}\left((1+\beta_E (P^{E})^2)({\bf X}^E-{\bf X}^{M})+\beta_E^{\prime}(({\bf X}^E-{\bf X}^M)\cdot{\bf P}^E){\bf P}^{E}\right),\label{E022}
\end{eqnarray}
and for the Moon
\begin{eqnarray}
\dot{{\bf X}}^M=\frac{{\bf P}^M}{m_M}\left(1+(\beta_M+\beta_M^{\prime})(P^{M})^2\right)-\frac{(2\beta_M-\beta_M^{\prime})+(2\beta_M+\beta_M^{\prime})\beta_M (P^{M})^2}{1+\beta_M (P^{M})^2}\times\nonumber\\\times\left(\frac{Gm_Mm_S [{\bf X}^M\times{\bf J}^M]}{R_{MS}^{3}}+\frac{Gm_Em_M[({\bf X}^M-{\bf X}^E)\times{\bf J}^M]}{R_{EM}^{3}}\right),\label{M020} \\
\dot{{\bf P}}^M=-\frac{Gm_Mm_S}{R_{MS}^{3}}\left((1+\beta_M (P^{M})^2){\bf X}^M+\beta_M^{\prime}({\bf X}^M\cdot\bf {P}^M){\bf P}^M\right)-\nonumber\\-\frac{Gm_Em_M}{R_{EM}^{3}}\left((1+\beta_M (P^{M})^2)({\bf X}^M-{\bf X}^{E})+\beta_M^{\prime}(({\bf X}^M-{\bf X}^E)\cdot{\bf P}^M){\bf P}^{M}\right),\label{M022}\nonumber\\
\end{eqnarray}
with ${\bf J}^M=[{\bf X}^{M}\times{\bf P}^M]$, ${\bf J}^E=[{\bf X}^E\times{\bf P}^E]$.
 Taking into account expressions for the effective parameters of deformation (\ref{efc1}), (\ref{efc2}), corresponding to  the Moon and the Earth, namely considering parameters $\beta_E$, $\beta^{\prime}_E$, $\beta_M$, $\beta^{\prime}_M$ to be defined as
 \begin{eqnarray}
\beta_E=\frac {\gamma^2}{m_E^2}, \ \ \beta^{\prime}_E=\frac{(\gamma^{\prime})^2}{m_E^2},\label{f1}\\
\beta_M=\frac {\gamma^2}{m_M^2}, \ \ \beta^{\prime}_M=\frac{(\gamma^{\prime})^2}{m_M^2},\label{f2}
\end{eqnarray}
from (\ref{E020})-(\ref{M022}) one obtains
\begin{eqnarray}
\dot{{\bf X}}^E={{\bf P}^{E\prime}}\left(1+(\gamma^2+(\gamma^{\prime})^2)(P^{E\prime})^2\right)+\nonumber\\+\frac{(2\gamma^2-(\gamma^{\prime})^2)+(2\gamma^2+(\gamma^{\prime})^2)\gamma^2 (P^{E\prime})^2}{1+\gamma^2 (P^{E\prime})^2}\times\left(\frac{Gm_S [{\bf X}^E\times[{\bf X}^{E}\times{\bf P}^{E\prime}]]}{R_{ES}^{3}}+\right.\nonumber\\\left.+\frac{Gm_M[({\bf X}^E-{\bf X}^M)\times[{\bf X}^{E}\times{\bf P}^{E\prime}]]}{R_{EM}^{3}}\right),\label{nE020}\\
\dot{{\bf P}}^{E\prime}=-\frac{Gm_S}{R_{ES}^{3}}\left((1+\gamma^2(P^{E\prime})^2){\bf X}^E+(\gamma^{\prime})^2({\bf X}^E\cdot{\bf P}^{E\prime}){\bf P}^{E\prime}\right)-\nonumber\\-\frac{Gm_M}{R_{EM}^{3}}\left((1+\gamma^2 (P^{E\prime})^2)({\bf X}^E-{\bf X}^{M})+(\gamma^{\prime})^2(({\bf X}^E-{\bf X}^M)\cdot{\bf P}^{E\prime}){\bf P}^{E\prime}\right),\label{nE022} \\
\dot{{\bf X}}^M={{\bf P}^{M\prime}}\left(1+(\gamma^2+(\gamma^{\prime})^2)(P^{M\prime})^2\right)+\nonumber\\+\frac{(2\gamma^2-(\gamma^{\prime})^2)+(2\gamma^2+(\gamma^{\prime})^2)\gamma^2 (P^{M\prime})^2}{1+\gamma^2 (P^{M\prime})^2}\times\left(\frac{Gm_S [{\bf X}^M\times[{\bf X}^{M}\times{\bf P}^{M\prime}]]}{R_{MS}^{3}}+\right.\nonumber\\\left.+\frac{Gm_E[({\bf X}^M-{\bf X}^E)\times[{\bf X}^{M}\times{\bf P}^{M\prime}]]}{R_{EM}^{3}}\right),\label{nM020}\\
\dot{{\bf P}}^{M\prime}=-\frac{Gm_S}{R_{MS}^{3}}\left((1+\gamma^2(P^{M\prime})^2){\bf X}^M+(\gamma^{\prime})^2({\bf X}^M\cdot{\bf P}^{M\prime}){\bf P}^{M\prime}\right)-\nonumber\\-\frac{Gm_E}{R_{EM}^{3}}\left((1+\gamma^2 (P^{M\prime})^2)({\bf X}^M-{\bf X}^{E})+(\gamma^{\prime})^2(({\bf X}^M-{\bf X}^E)\cdot{\bf P}^{M\prime}){\bf P}^{M\prime}\right), \label{nM022}
\end{eqnarray}
where ${\bf P}^{M\prime}={\bf P}^{M}/m_M$, ${\bf P}^{E\prime}={\bf P}^{E}/m_E$. Note that the obtained equations (\ref{nE020})-(\ref{nE022}) do not depend on $m_E$, therefore their solutions $X_i^{M}(t)$, $P_i^{M\prime}(t)$ do not depend on $m_M$ too. Similarly  $X_i^{E}(t)$, $P_i^{E\prime}(t)$ do not depend on  $m_E$. Therefore, the weak equivalence principle is recovered.

Let us calculate  accelerations of the Moon and the Earth toward the Sun when the distance from the Sun to the bodies is the same, $R_{MS}=R_{ES}=R$. We choose the origin of the frame of references to be at the Sun's center and consider the $X_1$ axis to be perpendicular to the ${\bf R}_{EM}$ and to pass through the middle of ${\bf R}_{EM}$.  Axis $X_2$ is considered to be parallel to ${\bf R}_{EM}$. So, $X^E_2=-X^M_2=R_{EM}/2$ and $X^E_1=X^M_1=R\sqrt{1-R^2_{EM}/4R^2}$, and $X^E_3=X^M_3=0$. Taking into account that $R_{EM}/R\sim10^{-3}$, one can write $X^E_1=X^M_1\simeq R$. From (\ref{nE020})-(\ref{nE022}) up to the first order in $\gamma^2$, $(\gamma^{\prime})^2$  one obtains the following expressions for the accelerations of the Earth and the Moon toward the Sun
\begin{eqnarray}
\ddot X^E_1=-\frac{Gm_S}{R^{2}}\left(1+2\gamma^2\upsilon^2_E\left(2-\frac{3R_{EM}^2}{4R^2}\right)+(\gamma^{\prime})^2\upsilon^2_E\frac{3R_{EM}^2}{4R^2}\right),\label{w}\\
\ddot X^M_1=-\frac{Gm_S}{R^{2}}\left[1+2\gamma^2\left(\upsilon^2_M\left(1+\frac{3R_{EM}^4}{4R^4}\right)+\upsilon^2_E\left(2-\frac{3R_{EM}^2}{4R^2}\right)+\right.\right.\nonumber\\ \left.\left.+\upsilon_E\upsilon_M\left(2-\frac{3R_{EM}^2}{8R^2}\right)\right)\right]-\frac{Gm_S}{R^{2}}(\gamma^{\prime})^2\left[\upsilon^2_M\left(2-\frac{3R_{EM}^2}{4R^2}\right)+\right.\nonumber\\\left.+\upsilon^2_E\frac{3R_{EM}^2}{4R^2}-\upsilon_E\upsilon_M\left(\frac{5R_{EM}}{2R}+\frac{m_ER^2}{m_S R_{EM}^2}-\frac{3R_{EM}^3}{8R^3}\right)\right].\label{ww}
\end{eqnarray}
To obtain (\ref{w}), (\ref{ww}) we take into account that in the ordinary case ($\beta_{E}=\beta^{\prime}_E=0$, $\beta_{M}=\beta^{\prime}_M=0$) $\dot{X}^E_1=0$,  $\dot{X}^E_{2}=\dot{X}^M_{2}=\upsilon_E$, $\dot{X}^M_1=\upsilon_M$, $\dot{X}^E_3=\dot{X}^E_3=0$,
with $\upsilon_E$, $\upsilon_M$ being the Earth and the Moon orbital velocities.
Using  (\ref{w}), (\ref{ww}) up to the first order in $\gamma^2$, $(\gamma^{\prime})^2$  one can find the Eotvos-parameter
\begin{eqnarray}
\frac{\Delta a}{a}=\frac{2(\ddot X^E_1-\ddot X^M_1)}{\ddot X_1^E+\ddot X_1^{M}}=2\gamma^2\left[\upsilon^2_M\left(1+\frac{3R_{EM}^4}{4R^4}\right)+\upsilon_E\upsilon_M\left(2-\frac{3R_{EM}^2}{8R^2}\right)\right]+\nonumber\\+
(\gamma^{\prime})^2\left[\upsilon^2_M\left(2-\frac{3R_{EM}^2}{4R^2}\right)-\upsilon_E\upsilon_M\left(\frac{5R_{EM}}{2R}+\frac{m_ER^2}{m_S R_{EM}^2}-\frac{3R_{EM}^3}{8R^3}\right)\right]\approx\nonumber\\
\approx\gamma^2\left(2\upsilon^2_M+4\upsilon_E\upsilon_M\right)+
(\gamma^{\prime})^2\left(2\upsilon^2_M-\upsilon_E\upsilon_M\left(\frac{5R_{EM}}{2R}+\frac{m_ER^2}{m_S R_{EM}^2}\right)\right)\label{eql}
\end{eqnarray}
Here we take into account  that $R_{EM}/R\sim10^{-3}$.
We would like to stress that from (\ref{eql}) one has that even if inertial and gravitational masses of the bodies are equal
(see (\ref{form12.5})) and even if equations of motion (\ref{nE020})-(\ref{nE022}), (\ref{nM020})-(\ref{nM022}) do not depend on the masses of the Earth and the Moon, respectively, the Eotvos-parameter is not equal to zero.
This effect is caused by the deformation. Namely, the accelerations of the Earth and the Moon toward the Sun depend on the velocities $\dot{X}^M_1$, $\dot{X}^E_1$ which are different for the bodies (Moon orbits the Earth with its orbital velocity $\dot{X}^M_1=\upsilon_M$, for the Earth this component of velocity is equal to zero $\dot{X}^E_1=0$). Therefore the accelerations of the Earth and the Moon toward the Sun are not the same.

The limit on the Eotvos-parameter for the Earth and the Moon was obtained on the basis of the results of the Lunar laser ranging experiment. It was found
\begin{eqnarray}
\frac{\Delta a}{a}=\frac{2(a_E-a_M)}{(a_E+a_M)}=(-0.8\pm 1.3)\cdot10^{-13}.\label{fe}
 \end{eqnarray}
 where $a_E$, $a_M$ are  free fall accelerations of the Earth and the Moon  toward the Sun in the case when the Earth and the Moon are at the same distance to the Sun \cite{Williams}.

 Assuming that the corrections to the Eotvos-parameter (\ref{eql}) caused by the deformation of commutation relations (\ref{0al})-(\ref{0al2}) are less than the limit  for ${\Delta a}/{a}$ obtained in \cite{Williams} one can write
  \begin{eqnarray}
\left|\gamma^2\left(2\upsilon^2_M+4\upsilon_E\upsilon_M\right)+
(\gamma^{\prime})^2\left(\upsilon^2_M-\upsilon_E\upsilon_M\left(\frac{5R_{EM}}{2R}+\frac{m_ER^2}{m_S R_{EM}^2}\right)\right)\right|\leq2.1\cdot10^{-13}\nonumber\\\label{in}
\end{eqnarray}
where $2.1\cdot10^{-13}$ is the largest value of (\ref{fe}).
 For estimation of the upper bound on the parameters of deformation it is sufficiently to consider the following inequalities
\begin{eqnarray}
\left|\gamma^2\left(2\upsilon^2_M+4\upsilon_E\upsilon_M\right)\right|\leq10^{-13},\\
\left|(\gamma^{\prime})^2\left(\upsilon^2_M-\upsilon_E\upsilon_M\left(\frac{5R_{EM}}{2R}+\frac{m_ER^2}{m_S R_{EM}^2}\right)\right)\right|\leq10^{-13},
\end{eqnarray}
 from which one obtains
$\gamma^2\leq 8\cdot10^{-22}s^2/m^2$, and $(\gamma^{\prime})^2\leq 8\cdot10^{-21}s^2/m^2$.
 Taking into account (\ref{f1}), (\ref{f2}), one has  $\beta_E\leq10^{-71}s^2/kg^2m^2$, $\beta^{\prime}_E\leq10^{-70}s^2/kg^2m^2$, $\beta_M\leq10^{-67}s^2/kg^2m^2$, $\beta^{\prime}_M\leq10^{-66}s^2/kg^2m^2$.
 Therefore for the minimal lengths corresponding to the Earth and the Moon one can find $l^{E}_{min}=\hbar\sqrt{\beta_E+\beta^{\prime}_E}\leq10^{-69}m$ and
$l^{M}_{min}=\hbar\sqrt{\beta_M+\beta^{\prime}_M}\leq10^{-67}m$.   For nucleons, using (\ref{c1}), (\ref{c2}),  one obtains
\begin{eqnarray}
l^{nuc}_{min}=\hbar\sqrt{\beta_{nuc}+\beta^{\prime}_{nuc}}\leq10^{-18}m\label{min}
\end{eqnarray}
Note that the obtained upper bounds for $l^{E}_{min}$, $l^{M}_{min}$ are many orders belove the Planck length. This is because of reduction of effective parameters of deformation corresponding to macroscopic bodies with respect to parameters corresponding to individual particles.  From (\ref{beff}), (\ref{beff1}) one has that for a composite system (a body) which consists of $N$ particles of the same masses and parameters $\beta$, $\beta^{\prime}$, one has $\tilde{\beta}=\beta/N^2$, $\tilde{\beta}^{\prime}=\beta^{\prime}/N^2$. So, the parameters $\tilde{\beta}$, $\tilde{\beta}^{\prime}$ corresponding to composite system reduce with increasing of number of particles which form it.

Putting  in (\ref{in})  $\gamma=0$ (which corresponds to $\beta=0$) one has that the upper bound for the minimal length in the Snyder space is of the order of $10^{-18}m$. For the deformed space with commuting coordinates (\ref{a1})-(\ref{a2}) in the case of $\beta^{\prime}=2\beta$ (or equivalently $(\gamma^{\prime})^2=2\gamma^2$) from (\ref{in}) one obtains upper bound of the minimal length of the same order $10^{-18}m$.
 The results are in agreement to that obtained in \cite{Tk2,Gnatenkoarxiv19} on the basis of studies of the Mercury motion.

\section{Conclusion}

A space with deformed Poisson brackets (\ref{pois1})-(\ref{pois3}) leading to the minimal length has been considered. We have studied a problem of description of motion of a body in the frame of algebra (\ref{pois1})-(\ref{pois3}).
It has been shown that in the space with minimal length the Poisson brackets for coordinates and momenta of the center-of-mass of a body do not reproduce relations of the deformed algebra (\ref{pois1})-(\ref{pois3}). We have found that  if  parameters of deformed algebra  (\ref{pois1})-(\ref{pois3}) depend on mass as (\ref{c1}), (\ref{c2}), the Poisson brackets for coordinates and momenta of the center-of-mass of a body correspond to relations of deformed algebra with effective parameters of deformation  (\ref{cmcc})-(\ref{cmcc2}).  We have also shown that in the deformed space there is a problem of violation of the property of independence of the kinetic energy of composition which can be solved  due to conditions (\ref{c1}), (\ref{c2}). In addition, we have found that if relations (\ref{c1}), (\ref{c2}) are satisfied  the equations of motion of a particle (a body) in the gravitational field do not depend on its  mass and composition, therefore the weak equivalence principle is recovered in the deformed space with minimal length. (\ref{pois1})-(\ref{pois3}).

 The Sun-Earth-Moon system has been studied in the frame of algebra (\ref{pois1})-(\ref{pois3}).  We have obtained that the  Eotvos-parameter for the Earth and the Moon is not equal to zero (\ref{eql})  even if inertial and gravitational masses of the bodies are equal and the trajectories of the bodies do not depend on their masses due to conditions (\ref{c1}), (\ref{c2}). This is because  deformation of Poisson brackets causes corrections to the accelerations of the Earth and the Moon toward the Sun. The corrections depend on the absolute values of velocities which are different for the bodies.

Comparing the obtained expression for the Eotvos-parameter in deformed space (\ref{eql}) with the  results of the Lunar Laser ranging experiment \cite{Williams} the upper bound for the minimal length has been estimated (\ref{min}). The result for the upper bound (\ref{min}) is in agreement to that obtained on the basis of studies of the Mercury motion \cite{Tk2,Gnatenkoarxiv19}.

\section*{Acknowledgements}

This work was partly supported by the Project $\Phi\Phi$-63Hp
(No. 0117U007190) from the Ministry of Education
and Science of Ukraine.

\end{document}